\documentclass[letterpaper,twocolumn,amsmath,amssymb,pre,aps,10pt]{revtex4-1}
\usepackage{graphicx} 
\usepackage{color}
\usepackage{nicefrac} 

\usepackage{xargs}                      
\usepackage[pdftex,dvipsnames]{xcolor}  
\usepackage[colorinlistoftodos,prependcaption,textsize=normalsize]{todonotes}
\usepackage{mdframed}
\usepackage{braket}

\begin{document}
\title{Flat-histogram method comparison on the 2D Ising model
}

\author{Jordan K. Pommerenck} \author{David Roundy}
\affiliation{Department of Physics, Oregon State University,
  Corvallis, OR 97331}

\begin{abstract}

We compare the convergence of several flat-histogram methods applied to the 2D
Ising model, including the recently introduced stochastic approximation with a
dynamic update factor (SAD) method. We compare this method with the Wang-Landau
(WL) method, the $1/t$ variant of the WL method, and standard stochastic
approximation Monte Carlo (SAMC). In addition, we consider a procedure WL
followed by a ``production run'' with fixed weights that refines the estimation
of the entropy. To our knowledge, this work is the first to test this approach
against other methods. We find that WL followed by a production run \emph{does}
converge to the true density of states, in contrast to pure WL. Three of the
methods converge robustly: SAD, $1/t$-WL, and WL followed by a production run.
Of these, SAD does not require \emph{a priori} knowledge of the energy range.
This work also shows that WL followed by a production run performs superior to
other forms of WL while ensuring both ergodicity and detailed balance.

\end{abstract}

\maketitle

\section{Introduction}
Flat-histogram Monte Carlo simulation algorithms calculate the thermodynamic
properties of various systems over a range of temperatures.  The first histogram
method used a single canonical Monte Carlo simulation to predict properties for
nearby temperatures~\cite{ferrenberg1988new}. While the method effectively
samples a narrow energy range, it proves computationally inefficient at sampling large energy ranges.
Multicanonical methods, introduced by Berg and Neuhaus, enabled flat-histogram
sampling which improved the exploration of configurational space and allowed the
simulation to overcome free-energy barriers~\cite{berg1991multicanonical, berg1992multicanonical}.
These works led to increase in the development of a variety of ``flat'' (or ``broad'') histogram
methods~\cite{penna1996broad, penna1998broad, swendsen1999transition,
wang2001determining, wang2001efficient} which could explore a wider range
of energies.  In addition to obtaining thermodynamic information for the entire
energy range for a single simulation, these approaches cannot be easily trapped
in a local energy minimum like a canonical simulation.

Wang and Landau introduced one of the most widely used flat-histogram
Monte Carlo algorithms that determined the density of states (DOS) for
a statistical system~\cite{wang2001determining,wang2001efficient}. For all of
its power, the method unfortunately requires \emph{a priori.} knowledge of several
user-defined parameters. Thus, for any given system under study, the user needs
to determine the ideal parameters in order to apply the method. The Wang-Landau
algorithm is also known to violate detailed balance (although only for brief
time intervals)~\cite{yan2003fast, shell2002generalization}. With the violation
of detailed balance, convergence of the algorithm is not guaranteed.

Because of the uncertainty of convergence for WL, many studies have been undertaken
to understand how the modification (or update) factor $\gamma$ impacts the
convergence~\cite{zhou2005understanding,lee2006convergence,
belardinelli2007wang}. Belardinelli and Pereyra showed that an update factor that decreases faster than $1/t$ leads to nonconvergence~\cite{belardinelli2007fast, belardinelli2007wang, belardinelli2008analysis, zhou2008optimal}, where $t$ corresponds to the number of moves. Schneider \emph{et al.} outline minor refinements algorithm including scaling the update factor with the number of energy bins~\cite{schneider2017convergence}.
These studies led to the
formation of the $1/t$-WL algorithm and have also led some researchers to follow
WL with a ``production run'' with fixed weights, in order to preserve ergodicity and
detailed balance~\cite{jayasri2005wang, mukhopadhyay2008monte}.
To our knowledge, this work is the first test of the convergence
properties of WL followed by a production run with comparison to other
methods.

Liang independently considered whether WL could be treated as a special case of
stochastic approximation whose convergence could be mathematically
proven~\cite{liang2006theory, liang2007stochastic}. In 2007, Liang \emph{et
al.}~\cite{liang2007stochastic} argued that WL can be considered a form of
stochastic approximation Monte Carlo (SAMC). Unlike WL, SAMC can guarantee
convergence (if certain conditions are met). Despite the added benefit of
guaranteed convergence, the method still has a system specific user-defined
variable. Such variables often create difficulty when applying Monte Carlo
methods across arbitrary systems.

Another challenge that flat-histogram methods face is that the convergence rate is
impacted by energy barriers and bottleneck which can make traversing the phase
space difficult. Nadler \emph{et al.} and Trebst \emph{et al.} systematically
examined optimized ensembles to address performance issues flat-histogram methods
face when confronted with hidden energy barriers~\cite{nadler2007generalized,
nadler2007dynamics, trebst2004optimizing}.
The approach involves using an ensemble that does \emph{not} result in a
flat histogram, but instead optimizes the rate of diffusion between low
energy and high energy states.
The goal of these methods is to make the statistical errors uniform.
Optimized ensemble methods typically begin by using a flat-histogram method
to get a first approximation for the weights, which means that any of the
methods
tested in this work could be used as a starting point for an optimized
ensemble simulation.


An approach to parallelizing flat-histogram Monte Carlo methods is
the replica-exchange approach, which was pioneered by parallel tempering
algorithms~\cite{geyer1995annealing, hukushima1996exchange,
hansmann1997parallel}.  Vogel \emph{et al.} adapted this approach to develop
the Replica Exchange Wang-Landau (REWL)~\cite{vogel2013rewl} approach.  This
approach for parallelization is sufficiently simple and general that it could
equally be applied to any of the methods explored in this paper.

Kim \emph{et al.} introduced Statistical Temperature Monte Carlo (STMC) and the
related Statistical Temperature Molecular Dynamics (STMD), an adaption of the
WL method that approximates the entropy as a piecewise linear function, which
improves convergence for systems with a continuously varying
energy~\cite{kim2006statistical, kim2007statistical}. STMC applied to WL
requires a temperature range be specified rather than an energy range. Kim
\emph{et al.} extended this work as Replica Exchange Statistical Temperature
Monte Carlo (RESTMC), which uses replica exchange of multiple overlapping STMC
simulations to improve convergence~\cite{kim2009replica}. Recently, Junghans
\emph{et al.} demonstrated a close connection between metadynamics, which was
introduced by Laio and Parinello~\cite{laio2002escaping}, and WL-based Monte
Carlo methods, with STMD forging the connection~\cite{junghans2014molecular}.

The SAD (stochastic approximation with a
dynamic $\gamma$) method as outlined by Pommerenck
et.al~\cite{pommerenck2020stochastic} is a special version of the SAMC algorithm
that dynamically chooses the modification factor rather than relying on
un-physical user-defined parameters. SAD shares the same convergence properties
with SAMC while replacing un-physical user-defined parameters with the
algorithms dynamic choice.

In this work, we compare the convergence properties of five flat-histogram
methods. We detail how each method is implemented and apply the family of
weight-based flat-histogram Monte Carlo methods (pure WL, WL followed by a
``production run'', $1/t$-WL, SAMC, and SAD) to the 2D Ising model.

\section{Ising Model}
The 2D Ising spin-lattice system is widely used as a testbed when
benchmarking or comparing Monte Carlo
methods~\cite{ferdinand1969bounded, wang1999transition, trebst2004optimizing, barash2019estimating}. The 2nd order
phase transition behavior and the ability to directly calculate the
exact solution for finite lattices~\cite{beale1996exact, haggkvist2004computation} make the
system sufficiently interesting for such theoretical comparisons. It is
also important to note that direct comparison of the other methods can
be made with WL as its original implementation was done on this
system~\cite{wang2001determining,wang2001efficient}.
We test the convergence of several flat-histogram methods
on the periodic 2D square lattice ferromagnetic Ising model with identical
nearest neighbor interactions~\cite{landau2004new} ($J_{ij} = J$).
\begin{align}
\mathcal{H} = -J \sum_{\braket{i,j}} \sigma_i \sigma_j - h \sum_i s_i
\end{align}
The $N\times N$ spin system can take on values of $\sigma_i = \pm 1$
for up or down spins respectively. In the absence of a magnetic field ($h =
0$), We can write the Hamiltonian as follows~\cite{onsager1944crystal,
kaufman1949crystal}:
\begin{align}
\mathcal{H} = - J\sum_{\braket{i,j}} \sigma_i \sigma_j
\end{align}
where the sum is over nearest neighbor spin sites. Beale showed that for finite
lattices the Density of States could directly be calculated from the partition
function~\cite{beale1996exact}
\begin{align}
Z = \sum_E g(E) e^{-{\beta E}}
\end{align}
where $g(E)$ is the multiplicity of the system which is proportional to
$D(E)$. We can compute the maximum deviation in the canonical specific heat capacity $c_V$ from the exact solution~\cite{schneider2017convergence, shakirov2018convergence,
barash2017control,barash2017gpu}:
\begin{align}
\epsilon \equiv \max_T \left|\left(c_V(T)_{\text{MC}} - c_V(T)_{\text{Beale}}\right)\right|
\end{align}
Computing the specific heat capacity $c_V$ presents a difficult challenge for
any Monte Carlo method due to fluctuations in the derivative of the
internal energy around the phase transition. The critical temperature $T_c = 2.
269T$ or $\beta= 0.441J$ comes directly from the Kramers-Wannier
duality~\cite{bhattacharjee1995fifty} and marks the transition from a
disordered to ordered magnetic state. Methods that accurately compute $c_V$
also by extension accurately compute the internal energy.

\section{Flat-histogram methods}\label{sec:histogram}
Flat-histogram methods compute the density of states $D(E)$ over a broad range
of energies by simulating each energy with equivalent accuracy. Flat-histogram
Monte Carlo methods propose randomly chosen ``moves'' which change the state of
the system and must satisfy detailed balance.  Each algorithm differs in how it
determines the probability of accepting a move and in what additional statistics
must be collected in order decide on that probability.

We describe several closely related flat-histogram methods which each rely on a
weight function $w(E)$ to determine $D(E)$.  For these algorithms, the
probability of accepting a move is given by
\begin{equation}
	\mathcal{P}(E_\text{old} \rightarrow E_\text{new})
	= \min\left[1,\frac{w(E_\text{old})}{w(E_\text{new})}\right]
\end{equation}
which biases the simulation in favor of energies with low weights. The result of
weights $w(E)$ that are proportional to $D(E)$ is an entirely flat-histogram. We
can relate the entropy to the weights in the microcanonical ensemble, since the
entropy is defined as $S(E) \equiv k_B\ln(D(E)) \approx \ln w(E)$.

\begin{figure}
\includegraphics[width=\columnwidth]{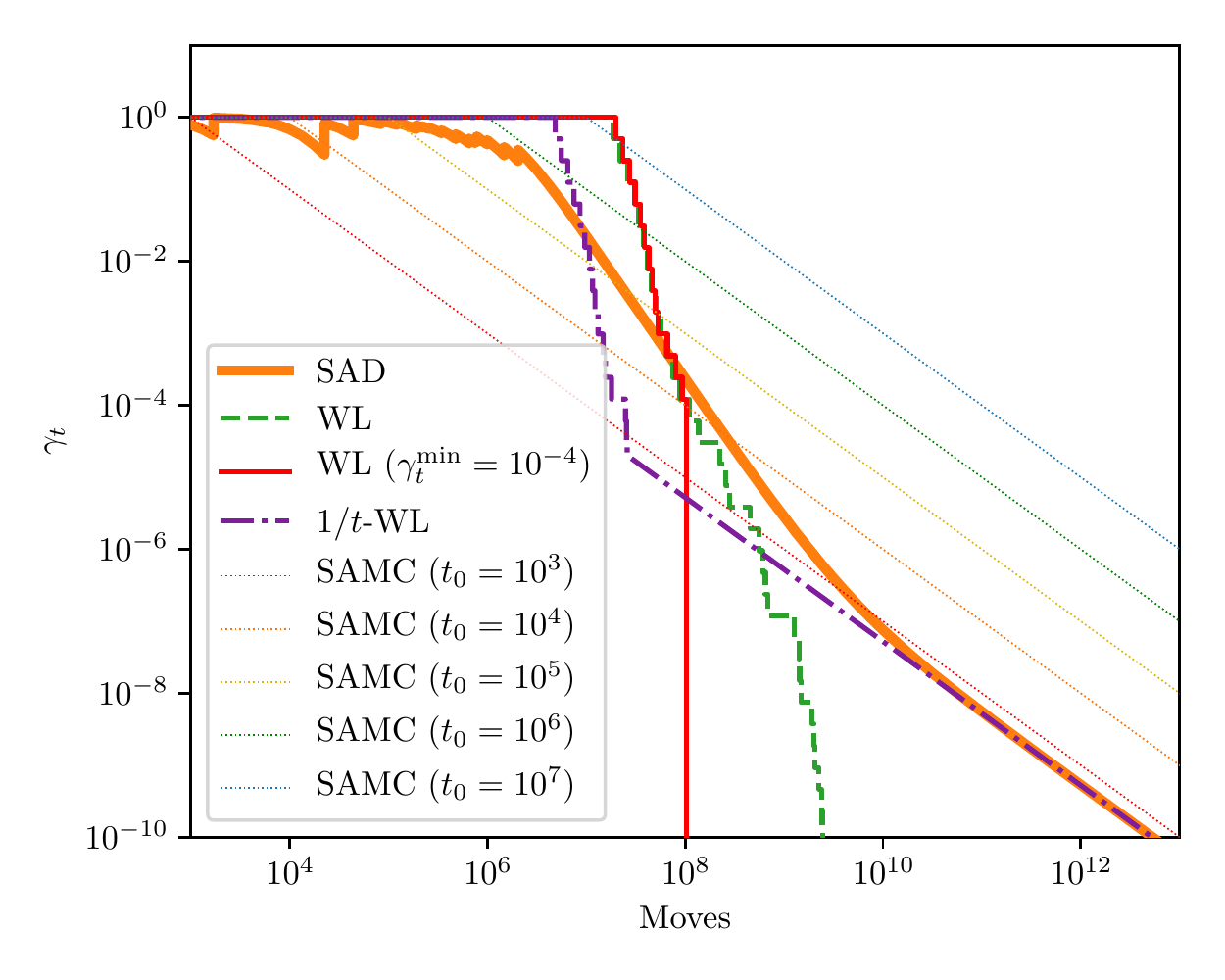}
  \caption{
  The update factor $\gamma_t$ versus the iteration number for the $N=32 \times 32$
  system.}
  \label{fig:N32-gamma}
\end{figure}

Flat-histogram methods employ a random walk in energy space to estimate $D(E)$.  Each method operates by continuously updating the weights at each
step of the simulation
\begin{equation}
	\ln{w_{t+1}(E)}=\ln{w_{t}(E)}
	+\gamma_t
\end{equation}
where $t$ is number of the current move, $\gamma(t)$ is a move-dependent update
factor, and $E$ is the current energy.  This update causes the random walk to
avoid frequent sampling of the same energies, leading to a rapid exploration
of energy space. Flat-histogram methods differ primarily in how they schedule
the decrease of $\gamma_t$.  Figure~\ref{fig:N32-gamma} shows several
flat-histogram methods each decreasing $\gamma_t$ as a function of moves.
Methods that decrease $\gamma_t$ too rapidly can fail to converge while methods
that decrease too slowly can take infinitely long to converge to the correct
$D(E)$.

The Wang-Landau algorithm~\cite{wang2001efficient,wang2001determining,
landau2014guide} explores energy space by setting
$\gamma_{t=0}^{\text{WL}}=1$, and then decreases $\gamma^{\text{WL}}$ in prescribed stages. An energy range of
interest must be specified~\cite{wang2001efficient, schulz2003avoiding,yan2003fast}, which often requires multiple simulations if unknown.
The number (``counts'') of moves ending at each energy are stored in a
histogram.  For a sufficiently flat energy histogram (typically user-specified to be 0.8), $\gamma^{\text{WL}}$ is decreased by a specified factor of $\frac12$ and the histogram is reset to zero. The entire process is repeated until $\gamma^{\text{WL}}$ reaches a desired cutoff.

The $1/t$-WL algorithm ensures convergence by preventing the $\gamma_t$ factor
from dropping below $N_S/t$~\cite{belardinelli2008analysis,
schneider2017convergence}. The method follows the standard WL algorithm with two
modifications.  Firstly, when each energy state has been visited once, the histogram is considered flat and $\gamma_t$ is
decreased by a factor of two. Secondly, when
$\gamma^{\text{WL}} < N_S/t$ at time $t_0$, the update factor becomes
$\gamma_t = N_S/t$ for the remainder of the simulation:
\begin{align}
  \gamma_t^{1/t\text{-WL}} = \begin{cases}
     \gamma^{\text{WL}}_t & \gamma^{\text{WL}}_t > \frac{N_S}{t} \\
     \frac{N_S}{t} & t \ge t_0
 \end{cases}
\end{align}
where $t$ is the number of moves, $\gamma^{\text{WL}}_t$ is the Wang-Landau update factor
at move $t$, and $N_S$ is the number of energy bins.

The WL method can be terminated after $\gamma$ reaches a specified
minimum $\gamma^{\min}$ followed by a \emph{production run} in which the
weights are held fixed~\cite{gross2018massively}. WL is used to
generate the weights resulting in a
flat-histogram~\cite{janke2017generalized}.  The entropy is then
computed (up to a constant) by adding the logarithm of the production
histogram to the logarithm of the weights.  The production run thus
satisfies detailed balance, and will ideally be ergodic; however, the
convergence of the simulation is still impacted by the choice of the
minimum $\gamma^{\min}$.

Another weight-based flat-histogram method is
the stochastic approximation Monte Carlo (SAMC) algorithm. SAMC has a simple
schedule by which the update factor $\gamma^{\text{SA}}_t$ is continuously
decreased~\cite{liang2007stochastic, werlich2015stochastic,
schneider2017convergence}.  The update factor is defined in the
original implementation~\cite{liang2007stochastic} in terms of an
arbitrary tunable parameter $t_0$,
\begin{align}
\gamma_{t}^{\text{SA}} =\frac{t_0}{\max(t_0,t)}
\end{align}
where as above $t$ is the number of moves that have been attempted.

The implementation of SAMC is extremely simplistic.
In addition, Liang has proven that the weights converge to the true
density of states~\cite{liang2006theory, liang2007stochastic,
liang2009improving} provided the update factor satisfies
\begin{align}
\sum_{t=1}^\infty \gamma_{t} = \infty \quad\textrm{and}\quad
\sum_{t=1}^\infty \gamma_{t}^\zeta < \infty
\end{align}
where $\zeta > 1$.  Unlike WL methods, the energy range need not be
known \emph{a priori.} and the convergence time depends only on the choice of
parameter $t_0$.
Unfortunately, $t_0$ can be difficult to chose in advance
for arbitrary systems.
Liang \emph{et al.} give a rule of thumb in
which $t_0$ is chosen in the range from $2N_S$ to $100N_S$ where $N_S$
is the number of energy bins~\cite{liang2007stochastic}.  Schneider
\emph{et al.} found and we confirm that for the Ising model this heuristic is helpful for small spin systems, but that larger systems require an even higher
$t_0$ value~\cite{schneider2017convergence}.

Pommerenck \emph{et al.} propose a refinement~\cite{pommerenck2020stochastic} to SAMC
where the update factor is determined dynamically rather than by the user.
Stochastic approximation with a dynamic $\gamma$ (SAD) requires the user to
provide a minimum temperature of interest $T_{\min}$. This is analogous to WL
requiring \emph{a priori.} an energy range of interest; however, this is almost always easier to
identify and is more physical than the SAMC parameter $t_0$. The
update factor is given by:
\begin{align}
  \gamma_{t}^{\text{SAD}} =
     \frac{
       \frac{E_{H}-E_{L}}{T_{\text{min}}} + \frac{t}{t_L}
     }{
       \frac{E_{H}-E_{L}}{T_{\text{min}}} + \frac{t}{N_S}\frac{t}{t_L}
     }
\end{align}
where $E_H$ and $E_L$ are the current estimates for the highest and
lowest interesting energies and $t_L$ is the last time at which an
energy in the range of interest is encountered.
SAD only explores the energy range of interest as specified by the minimum
temperature of interest $T_{\min} < T < \infty$. During the simulation the two
energies $E_H$ and $E_L$, are refined such that the range of energies are
conservatively estimated. The weights are calculated for each energy region according to the original prescription.
\begin{enumerate}
\item {$E < E_L$:} $w(E>E_H) = w(E_H)$
\item {$E_L < E < E_H$:} moves are handled the same as other weight-based
methods that are mentioned
\item {$E > E_H$:} $w(E<E_L) = w(E_L)e^{-\frac{E_L-E}{T_{\min}}}$
\end{enumerate}
Each time the simulation changes the value of $E_H$ or $E_L$, the weights
within the new portion of the interesting energy range are updated.

\section{Results}
We test the algorithms on two different system sizes of the 2D Ising model.
The first is a smaller simulation with a lattice size of $N = 32 \times 32$ and
the second has a lattice size of $N = 128 \times 128$. The SAD method explores 
the energy space of each system using a minimum reduced temperature of $T_{\text{min}} = 1$. All simulations calculate the minimum important energy $E_{\min}$
and maximum entropy energy $E_{\max}$ (with the exception of the WL methods
where both of these parameters are needed \emph{a priori.}).


\begin{figure}
  \includegraphics[width=\columnwidth]{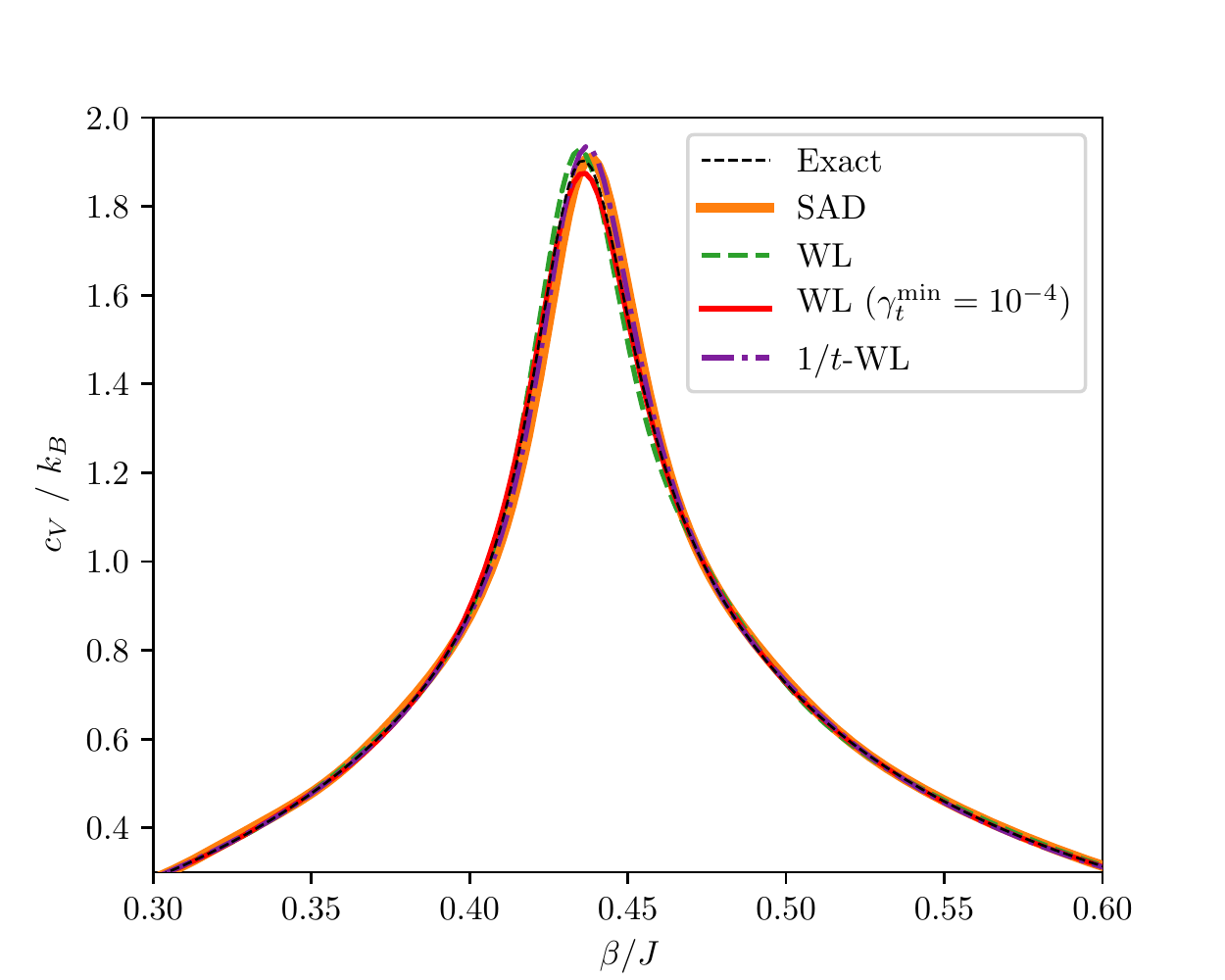}
  \caption{
    The specific heat capacity versus the reciprocal temperature $\beta$ for the $N=32 \times 32$ system for each histogram method at $10^{9}$ moves.}
  \label{fig:N32-cv}
\end{figure}

\subsection{The 32 $\times$ 32 Ising model}
Figure~\ref{fig:N32-gamma} shows the update factor $\gamma_t$ for each
of the flat-histogram methods. All of the update factors initially
start at $\gamma_{t=0} = 1$. SAD dynamically updates $\gamma_t$
throughout the simulation. After about $10^{10}$ moves,
$\gamma^{\text{SAD}}_t$ proceeds as $1/t$. We show
$\gamma^{\text{WL}}_t$ for WL both with and without a production
run. The WL production run begins after $\gamma^{\text{WL}}_t$ has reached
$10^{-4}$. The update factor for $1/t$-WL decreases similarly to WL
before finding all the energy states and switching to $1/t$. All of
the SAMC update factors equal 1 until the number of moves is equal to
$t_0$ at which point they decrease as $1/t$.

All of the methods, except for SAMC, use the same single random number seed and
give a reasonable approximation for the heat capacity peak resulting from the
phase transition after only 10$^9$ moves. Figure~\ref{fig:N32-cv} shows the
specific heat capacity vs. the reciprocal temperature $\beta$ at $10^{9}$ moves.
A temperature range of $1.5k_B$ to $5k_B$ is chosen to highlight
the phase transition at the critical temperature $T_c$.

\begin{figure}
  \includegraphics[width=\columnwidth]{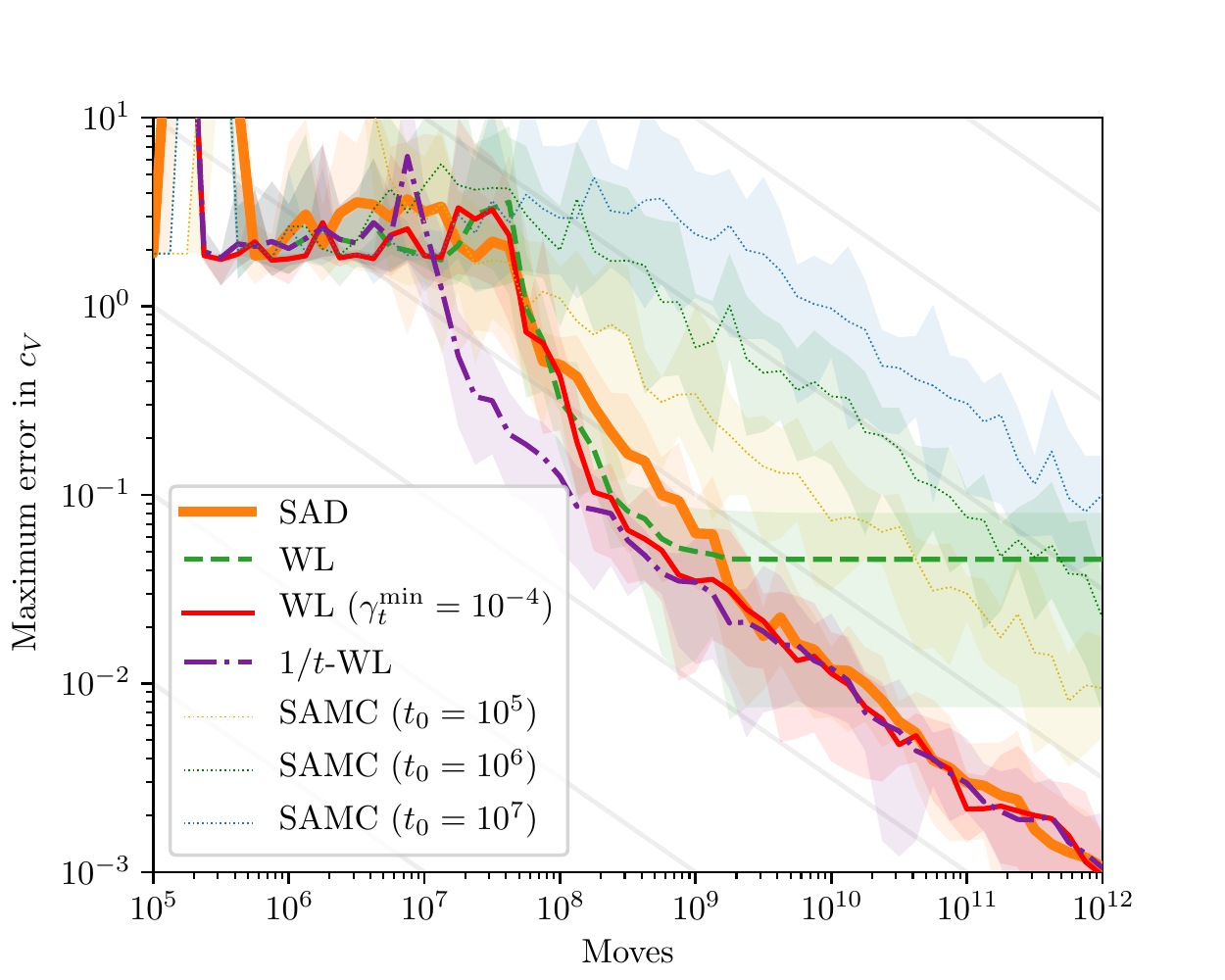}
  \caption{The maximum error in the specific heat capacity for each method for the $N=32 \times 32$ and $T_{\min} = 1$ as a function of number of iterations run.  The maximum error is averaged over 8 independent simulations, and the best and worst simulations for each method are shown as a semi-transparent shaded area.}
  \label{fig:N32-cv-error}
\end{figure}

Figure~\ref{fig:N32-cv-error} shows the maximum error in the heat capacity as a
function of time for this system. The solid/dashed lines represent the average
of the maximum value of the error in the specific heat capacity $c_V$ averaged
over eight simulations using different random number seeds. The range of maximum
errors for each simulation is shown as a shaded region. By
the time $10^8$ moves have been made all but the WL simulation have begun to
converge as $1/\sqrt{t}$. We then see the WL error saturate around $10^9$ moves.

\begin{figure}
\includegraphics[width=\columnwidth]{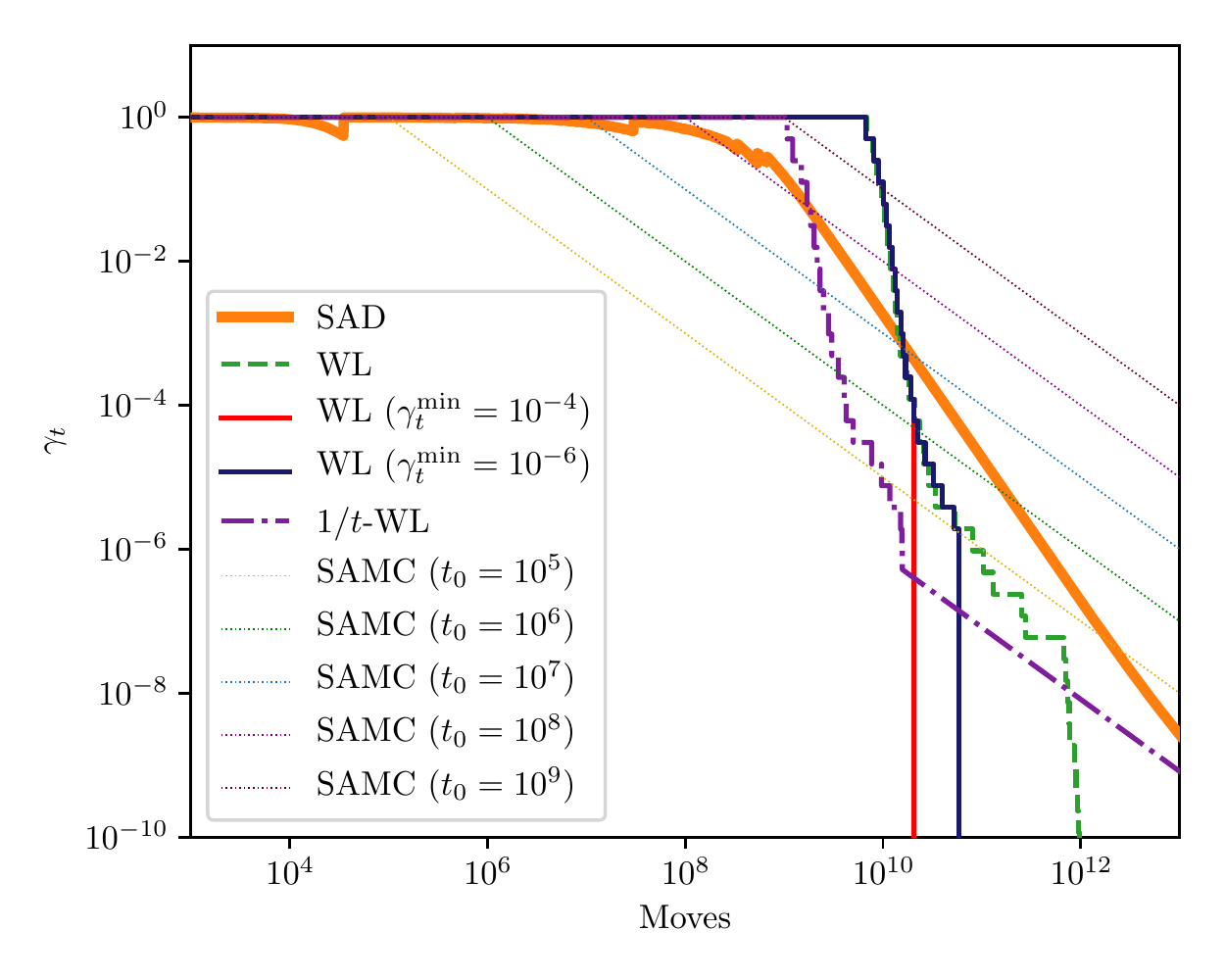}
  \caption{
  The update factor $\gamma_t$ versus the iteration number for the $N=128 \times 128$
  system.}
  \label{fig:N128-gamma}
\end{figure}

\begin{figure}
  \includegraphics[width=\columnwidth]{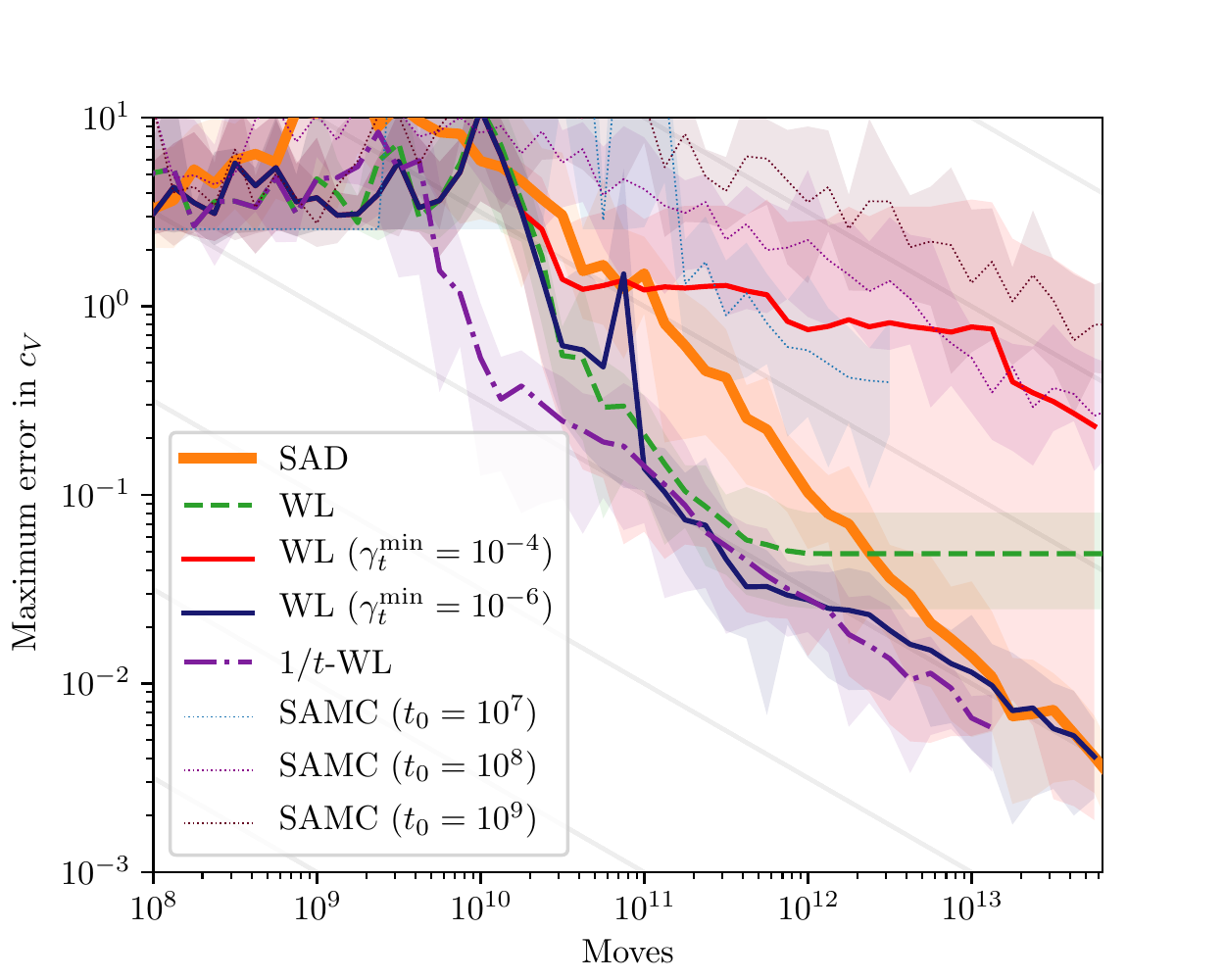}
  \caption{The maximum error in the specific heat capacity for each method for the $N=128 \times 128$ and $T_{\min} = 1$ as a function of number of iterations run.  The maximum error is averaged over 8 independent simulations, and the best and worst simulations for each method are shown as a semi-transparent shaded area.}
  \label{fig:N128-cv-error}
\end{figure}


\subsection{128 $\times$ 128 Ising system}
For the larger 2D Ising system, the update factors initially
start at the same $\gamma_{t=0} = 1$; however, all of the methods take longer
to proceed as $1/t$ (with SAD taking $10^{13}$ moves). Fig.~\ref{fig:N128-gamma} shows the update factor $\gamma_t$ for each of the 
flat-histogram methods. 
We implement two WL simulations followed by a production run each beginning
after $\gamma^{\text{WL}}_t$ has reached $10^{-4}$ and $10^{-6}$ respectively.
Fig.~\ref{fig:N128-cv-error} shows the maximum error in the heat capacity as a 
function of time for this system. The solid/dashed lines represent the maximum
value of the error in the specific heat capacity $c_V$ averaged
over eight simulations using different random number seeds. The range of maximum
errors for each simulation is shown as a shaded region. By
the time $10^{13}$ moves have been made all but the WL simulation have begun to
converge as $1/\sqrt{t}$. We then see the WL error saturate around $10^{12}$
moves.  The WL simulation followed by a production run when
$\gamma^{\text{WL}}_t$ reached $10^{-4}$ sometimes dramatically failed to
converge, although five of the eight random seeds converged very
nicely.  This highlights a risk taken when setting $\gamma^{\min}$.
If the weights are insufficiently converged, the production run will
fail to explore all energies, in this case, three of the simulations became ``stuck'' at low energies. With the smaller value of
$\gamma^{\min}=10^{-6}$, the method consistently and efficiently converged.

Figure~\ref{fig:N128-cv-error} shows that the SAD algorithm converges
significantly more slowly than the converging WL methods.
The SAD algorithm on average takes around
$10^9$ moves to identify as important all 8192
negative energy states, which is around twice as long as
the number of moves that
the WL methods require in order to explore all the states in this
energy range.  The main difference between the convergence of these
methods is that the WL-based methods decrease $\gamma_t$ far more rapidly,
which leads to more rapid convergence.  The Wang-Landau approaches can
get away with this because the range of energies of interest is given
as an input rather than an output, allowing a more aggressive schedule
of reduction of $\gamma_t$.  This aggressive behavior is precisely what
requires that Wang-Landau methods be followed by some correction stage
(either $1/t$ or a production run) in order to correct residual errors.

\section{Conclusion}
We find that SAD, $1/t$-WL, and WL followed by a production run (with an
adequately small $\gamma^{\min}$) demonstrate excellent and robust
convergence. They all converge more rapidly than SAMC, and unlike pure
WL do not suffer from error saturation. We find that for larger Ising
systems SAD reduces the update factor more slowly (and conservatively)
than $1/t$-WL and WL followed by a production run. This means that SAD
will take proportionately more moves to converge to the same value as
$1/t$-WL as system size is increased.  While the WL methods are
are given the energy range
\emph{a priori}, rather than a temperature range of interest such as
SAD requires, we find that the SAD histogram counts for energies outside the range of interest are negligible. For the general case in
which the energy range of the system is not known and where a range of
desired temperatures is known, the SAD method is
considerably more convenient, and quite possibly more efficient than a
process involving multiple simulations to determine an energy range
of interest. This work also dramatically demonstrates that WL followed
by a production run performs extremely well and is preferable to pure
WL for ensuring both ergodicity and detailed balance. While this has long
been thought to be the case and that a multi-canonical run where WL is used to
determine the parameters is the only way WL should be used, this research
represents the first detailed comparison among all of these
flat-histogram methods.

\section{Acknowledgments}
We wish to thank Johannes Zierenberg for helpful discussions regarding WL followed by a production run and for insights into properties of a production run at different stages of convergence.

\bibliography{ising} 

\end{document}